\documentclass[aps,pra,amsmath,amssymb,graphicx,longbibliography, twocolumn, superscriptaddress]{revtex4-2}

\usepackage[utf8]{inputenc}
\usepackage{bm}
\usepackage{graphicx}
\usepackage{epstopdf}
\usepackage{wrapfig}
\usepackage{array}
\usepackage{listings}
\usepackage{dsfont}
\usepackage[para,online,flushleft]{threeparttablex}
\usepackage{booktabs,dcolumn}
\usepackage{color}
\usepackage{textpos}
\usepackage{booktabs}
\usepackage{multirow,bigdelim}
\usepackage{float}
\usepackage{upgreek} 
\usepackage{xcolor}
\usepackage{comment}
\usepackage{physics}
\usepackage{braket}
\usepackage{environ}
\usepackage{bbm}
\usepackage{xspace}
\usepackage{isotope}
\usepackage{tabularx, makecell}

\definecolor{steelblue}{RGB}{0, 51, 97}

\usepackage[unicode=true,
 bookmarks=false,
 breaklinks=false, colorlinks=true]
 {hyperref}
\hypersetup{
  linkcolor = steelblue,
  citecolor = steelblue,
  urlcolor = steelblue,
}

\newcolumntype{C}[1]{>{\centering\arraybackslash}p{#1}}

\begin{document}

\title{Improved Rodeo Algorithm Performance for Spectral Functions and State Preparation}

\author{Matthew Patkowski}
\affiliation{Department of Physics, University of Colorado Boulder, Boulder CO 80302}

\author{Onat Ayyildiz}
\affiliation{Department of Physics, University of Michigan, Ann Arbor MI 48109}
\affiliation{Facility for Rare Isotope Beams, Michigan State University, East Lansing MI 48824}
\affiliation{Department of Physics \& Astronomy, Michigan State University, East Lansing MI 48824}

\author{Katherine Hunt}
\affiliation{Department of Chemistry, Michigan State University, East Lansing MI 48824}

\author{Nathan Jansen}
\affiliation{Department of Chemistry, Michigan State University, East Lansing MI 48824}

\author{Dean Lee}
\affiliation{Facility for Rare Isotope Beams, Michigan State University, East Lansing MI 48824}
\affiliation{Department of Physics \& Astronomy, Michigan State University, East Lansing MI 48824}

\begin{abstract}
The rodeo algorithm is a quantum computing method for computing the energy spectrum of a Hamiltonian and preparing its energy eigenstates.  We discuss how to improve the performance of the rodeo algorithm for each of these two applications. In particular, we demonstrate that using a geometric series of time samples offers a near-optimal optimization space for a given total runtime by studying the Rodeo Algorithm performance on a model Hamiltonian representative of gapped many-body quantum systems. Analytics explain the performance of this time sampling and the conditions for it to maintain the established exponential performance of the Rodeo Algorithm. We finally demonstrate this sampling protocol on various physical Hamiltonians, showing its practical applicability. Our results suggest that geometric series of times provide a practical, near-optimal, and robust time-sampling strategy for quantum state preparation with the Rodeo Algorithm across varied Hamiltonians without requiring model-specific fine-tuning.
\end{abstract}

\maketitle
\section{Introduction}

Quantum computation offers a promising possibility for the simulation of complex quantum problems, including condensed matter systems, quantum many-body physics, and quantum chemistry, among others \cite{georgescu2014quantum, greiner2002quantum, aspuru2005simulated}. A crucial step in quantum simulation is quantum state preparation, for which the traditional approaches are techniques including adiabatic state preparation and quantum phase estimation \cite{kitaev1995quantum, farhi2000quantum}. A recent approach termed the Rodeo Algorithm (RA) offers exponential convergence compared to these techniques \cite{Choi:2020pdg}.
However, while the RA offers exponential suppression of unwanted states, its efficiency depends on the choice of time sampling, which is often stochastic, leading to varying performance shot-to-shot.
The optimization of quantum algorithms, including the RA, is a crucial step toward realizing scalable and practical quantum simulation. In particular, minimizing the runtime and resources required by a quantum algorithm mitigates the effects of finite coherence times and other experimental noise sources.

Improvements to the RA performance include a reduction in quantum gate complexity \cite{bee2025controlled} and improved non-stochastic time sampling protocols in \cite{cohen2023optimizing}.
In particular, the latter study demonstrated that a geometric sequence of common ratio 1/2 offers good performance, and its non-stochastic nature allows for more reliable performance.
Similar projection algorithms have also found the same time series of $t_i = t_{i-1}/2$ to perform well \cite{stetcu2023projection}.
In this work, we expand the analysis of the performance of the RA through the analysis of two metrics that quantify the RA performance. Building on the intuition from \cite{cohen2023optimizing} and numerical evidence, we study a particular set of times we term ``generalized superiterations'' which are a geometric sequence of times with a common ratio $\alpha ^{-1}$, which we numerically demonstrate to be a near-optimal solution, and exhibit exponential scaling given proper optimization over the parameter space.
We offer analytical arguments to explain the performance of these time sampling schemes, as well as demonstrate their performance on physical Hamiltonians.
We emphasize that optimizing a single parameter $\alpha$ is computationally trivial compared to optimizing an entire vector of $N$ time samples. This ``single-parameter variation" allows for efficient run-time adjustment without requiring expensive, model-specific fine-tuning.

The structure of the manuscript is as follows. In Section~\ref{sec:metrics}, we introduce two metrics to quantify the RA performance. In Section~\ref{sec:modelham}, we introduce a simple Hamiltonian representative of gapped systems and demonstrate numerical evidence that the generalized superiteration ansatz contains an optimization minimum that is near-optimal. In Section~\ref{sec:asymptotics}, we study the RA performance metrics analytically, demonstrating that the generalized superiterations applied naively perform with a power-law increase in fidelity, but can be improved to recover the exponential scaling with a carefully-chosen $\alpha$. In this section, we also demonstrate connections between the RA performance with generalized superiterations and the number-theoretical properties of $\alpha$. Finally, in Section~\ref{sec:physicalham} we demonstrate the practical applicability of the generalized superiterations across a range of physical Hamiltonians.

\section{Quantifying the Rodeo Algorithm Performance}
\label{sec:metrics}

The RA is a quantum algorithm that takes an input state $\ket{\psi}$ and purifies it toward an eigenstate of a given Hamiltonian $H$ with given ``target" energy $E_t$. Controlled time evolution by the Hamiltonian is evolved for the $N$ time samples $\{t_n\}$. Conditioned on the successful application of the algorithm, which is post-selected based on the values of ancilla qubits, the post-RA state can be written
\begin{equation}
\label{eq:postrastateexpression}
    \ket{\psi '}= \prod _{n=1}^N \frac{ \mathds 1+ \exp [i(E_t -   H)t_n] }2 \ket{\psi }
\end{equation}
Projecting Eq.~\eqref{eq:postrastateexpression} into the eigenbasis of $ H$, we obtain
\begin{equation}
\label{eq:postrastateexpressionineigenbasisofH}
    \xi '(E) =  \xi (E) \prod _{n=1}^N\frac{ 1+ \exp [i(E_t -E)t_n]}2
\end{equation}
where we introduced the pre- and post-RA spectral functions
$\xi (E) \equiv \braket{ E|\psi }$ and $\xi ' (E) \equiv \braket{E | \psi '}$.
As written, the transformation in Eq.~\eqref{eq:postrastateexpression} is non-unitary, which follows from the gate-based implementation, where the computational system is entangled with ancillary qubits and the output state is post-selected based on the measurement outcomes of the ancilla qubits \cite{Choi:2020pdg}. As seen from Eq.~\eqref{eq:postrastateexpressionineigenbasisofH}, the RA preserves the norm of $\ket {E_t}$, while reducing the amplitude of all other eigenstates. Eigenstates of energy $E$ with $(E_t - E)t_n \approx (2k+1)\pi $ where $k\in \mathbb Z $ are most strongly suppressed, while eigenstates for which $(E_t-E)t_n \approx 2k \pi$ remain largely unsuppressed. We hence define the ``characteristic time" $T_0 = \pi \Delta _{\rm min}^{-1}$ which is the time for perfect suppression of the closest eigenstate $E$ to $E_t$ with $|E_t - E |=\Delta _{\rm min}$.

We introduce two metrics to quantify the performance of the RA. One is the success probability $p(E,t)= \cos ^2 \left[\left(E-E_t\right) \tfrac t2\right]$ which, by Eq.~\eqref{eq:postrastateexpressionineigenbasisofH}, is the reduction in the norm of eigenstate $\ket E$ for an RA application with single time sample $t$. Evidently, minimizing the success probability of a given eigenstate results in the strongest suppression of that eigenstate. We comment that minimizing the success probability minimizes the chance of RA success in the post-selection process, and in the ideal case the success probability for all eigenstates other than the target will go to zero, leaving a total success probability $|\braket{\psi |E_t}|^2.$ However, the strategy we suggest is using other methods, such as a preconditioning of the input state or lattice construction methods, as developed in \cite{patkowski2025high, bogner2025quantum}, to maximize the overlap between the input state and the target state before using the RA as a fast purification step. In other words, minimizing the success probability gives the best suppression performance given that one accounts for repetitions due to post-selection.

The second metric is the residual spectral norm (RSN), defined as
\begin{equation}
\label{eq:residualspectralnorm}
\zeta \equiv \int _{\mathbb R \setminus \{E_t\}} \mathrm dE \, |\xi '(E)|^2 = \int_{\mathbb R \setminus \{E_t\}}  \mathrm dE \, |\xi(E)|^2 \prod _ {n=1}^N p(E,t_n),
\end{equation} which is minimized for optimal suppression. Throughout this work, $\zeta$ should be interpreted as the total weight of the undesired eigenstates before post-selection normalization. In this way, minimizing $\zeta$ is equivalent to maximizing post-selected fidelity with $\ket{E_t}$.

\section{Model Hamiltonian and Assumptions}
\label{sec:modelham}
As implied by Eq.~\eqref{eq:postrastateexpressionineigenbasisofH}, given one has access to the initial spectral function $\xi (E)$ and the full spectrum of the Hamiltonian, in $N= \dim  H$ steps one could perfectly suppress all undesired eigenstates, perfectly suppressing $\zeta \to 0$. In practice, this is impossible, however, due to (1) imperfect knowledge of the spectrum of $ H$ and (2) finite resources limiting $N$. Problem (1) can be mostly alleviated by utilizing the energy sweeping functionality of the RA \cite{Choi:2020pdg}, allowing precise and efficient access to the spectrum of $ H$. Contrastingly, problem (2) is unavoidable, since the Hilbert space dimension generally grows exponentially in system size, rendering any approach that targets individual eigenstates with separate time steps prohibitively inefficient.
We therefore assume only a coarse-grained understanding of the spectrum without precise knowledge of $\xi (E)$ and eigenvalues of $ H$.

To introduce a convenient trial model, we make several simplifying assumptions about the problem setup.
We consider a Hamiltonian whose ground state of interest has energy $E_0 = E_t$, with a continuous spectrum $E\in [\Delta_{\text{min}} +E_0, \Delta_{\text{max}} + E_0],$ where we have a finite excitation gap $\Delta_{\text{min}}  > 0.$ We shift the spectrum such that $E_t = E_0 = 0$. Models with a gapped ground state and a continuum of excitations are standard in quantum many-body physics and condensed-matter systems \cite{bloch2008many}.
Additionally, we assume that $\xi (E)$ is a constant function of $E$, which assumes we lack spectral knowledge and do not perform preconditioning steps before starting the RA. By introducing a continuous band of energies, we eliminate artificial dips in the RSN that would result from perfect suppression of discrete eigenstates if we had not taken this limit.

\begin{figure}[t]
    \centering
    \includegraphics[width=\linewidth]{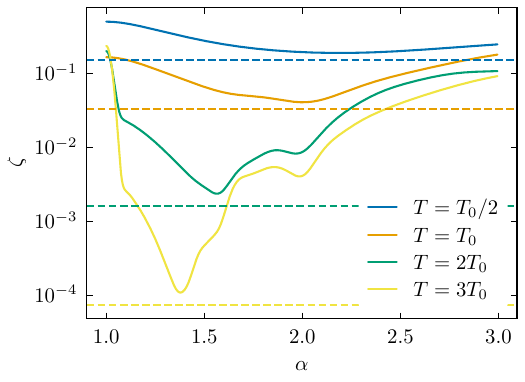}
    \caption{Examples results shown for $N=10$, $\Delta _{\text{min}} = 0.1$, $\Delta _{\text{max}} = 1$. Solid lines correspond to time samples in the generalized superiteration subspace, while dashed lines correspond to the global optimum across all time samples with $N\leq 10$ and total time less than or equal to the indicated constraint.}
    \label{fig:example_gensuper_results}
\end{figure}

Under these assumptions, the initial spectral function is pulled out of the integral in Eq.~\eqref{eq:residualspectralnorm} and further simplifies to the expression in Eq.~\eqref{eq:rsn_constant_overlap_continuous_spectrum}, as derived in App.~\ref{app:suppressionderivation}.
The expression is intractable to minimize analytically due to the complicated landscape of the sum of sinc functions across the $2^{2N}$ terms. Still, we may numerically minimize for modest $N \lesssim 15$ to gain insight into optimal time sampling choices. We present numerical optimization results in Table~\ref{tab:continuousspecoptres}. The form of the optimal time samples suggests that a geometric series of time samples is a well-suited ansatz. In Fig.~\ref{fig:example_gensuper_results} the solid lines represent time samples restricted to the space of a geometric series with common ratio $\alpha ^{-1}$. The corresponding dashed lines represent the global minimum without restricting the relation between time samples, only restricting the total time to be less than the indicated time. We see that for this model Hamiltonian, the optimization subspace corresponding to geometric series of times (termed ``generalized superiterations"), when optimized over $\alpha$, is able to reach near the global minimum of times. We note that for times similar to and less than the characteristic time, the optimal $\alpha$ is 2. On the other hand, when we have the resources to permit the total time to be longer than the characteristic time, we see a shift of the optimal $\alpha$ to values less than 2, moving toward 1 as the total time is increased. We note that the dips in the RSN about the optimal $\alpha$ are not fine-tuned in $\alpha$, suggesting this sampling of times is resilient to noise sources that may change the exact time evolution we implement, as well as Trotterization error.

\begin{table}[t]
\centering
\begin{tabularx}{\linewidth}{|c|X|c|}
\hline
Time Constraint & $\{t_n\}$ & $\zeta$ \\
\hline
$T \le T_0/2$ &
\makecell[l]{$(0.449, 4.956, 10.302)$} &
$0.153$ \\
\hline
$T \le T_0$ &
\makecell[l]{$(3.463, 4.300, 8.058, 15.596)$} &
$0.0335$ \\
\hline
$T \le 2T_0$ &
\makecell[l]{$(3.382, 4.118, 6.312, 10.303,$ \\ $14.929, 23.788)$} &
$0.00161$ \\
\hline
$T \le 3T_0$ &
\makecell[l]{$(3.323, 4.210, 5.738,$ \\$ 7.843,  11.097, 14.557,$\\ $19.829, 27.650)$} &
$7.42\times 10^{-5}$ \\
\hline
\end{tabularx}
\caption{Optimization results for $\Delta _{\rm min} = 0.1,\Delta _{\rm max} = 1,$ and $N=10$. Optimization performed via the Python cma package, see \cite{PatkowskiRODEOCode} for code. Time samples less than $10^{-6}$ are discarded. We define the total time $T\equiv \sum _n t_n.$}
\label{tab:continuousspecoptres}
\end{table}

\section{Asymptotics of Success Probability}
\label{sec:asymptotics}
In practice, effective suppression of undesired eigenstates requires time evolution over durations on the order of several $T_0$, and potentially longer. Accordingly, we analyze the suppression properties of the RA in the limit $T\to \infty$. We first define the product function
\begin{equation}
    C(\alpha ,\theta, N) = \prod _{n=1}^N \cos ^2 \left[\frac{(\alpha -1) \theta}{\alpha ^n}\right]
\end{equation} for real $\theta$ and $\alpha >1$. We note that when $\alpha = 2$ and $\theta = (E-E_t) \tfrac t2$, the above equation corresponds to the product of success probabilities for a given eigenstate $E$ with a geometric series with initial value $t$ and common ratio $1/2$, corresponding to the superiteration time samples used in \cite{cohen2023optimizing}. For other values of $\alpha$, this corresponds to the product of success probabilities for a given eigenstate $E$ with generalized superiterations with common ratio $\alpha ^{-1}$ and initial value $t_1 = (\alpha -1) t$. By understanding the asymptotics of the product function, we can bound the RSN $\zeta$ and thus the integrated performance over a Hamiltonian's spectrum.

We first study the behavior of the product function
\begin{equation}
    C\left(1+\frac b\theta , \theta,N\right) = \prod _{n=1}^N \cos ^2 \left(\frac{b}{\left(1+\tfrac b\theta\right)^n}\right)
    \label{eq:appr1above}
\end{equation}
in the limit $\theta \to \infty$ with $b>0$ fixed, corresponding to the regime where $\alpha$ approaches 1 from the above.
In this limit, Eq.~\eqref{eq:appr1above} takes the form
\begin{equation}
    \lim _{\theta \to \infty}C\left(1+\frac b\theta , \theta,N\right) = \exp \left[2N\log \cos b + \mathcal O(N^2/\theta)\right],
\end{equation}
demonstrating exponential suppression of the eigenstate for large $N$ kept small relative to $\theta$.

Next, we examine the asymptotic properties of the success probability product function at fixed $\alpha$ and $N\to \infty$. Define $\lambda = 1/\alpha \in (0,1)$ and rescale $\tilde \theta = \theta (\alpha - 1)$. The classical Bernoulli convolution with parameter $\lambda$ is the probability measure
\begin{equation}
    \nu _\lambda = \lim _{N\to \infty} \sum _{\varepsilon \in \{-1,1\}^N} \frac{1}{2^N} \delta \left(x-\sum _{n=1}^N \varepsilon _n \lambda ^n\right)
\end{equation}
where the limit is in the weak limit sense. Its Fourier transform is
\begin{equation}
    \widehat \nu _\lambda = \prod _{n=1}^\infty \cos (\lambda ^n t).
\end{equation}
This establishes the exact correspondence
\begin{equation}
    \lim _{N\to \infty}C(\alpha, \theta,N) = C(\alpha,\theta)= \left |\widehat \nu _{1/\alpha} \left[(\alpha -1)\theta\right] \right| ^2,
\end{equation}
and hence has identical asymptotic decay (or lack thereof).
Using $\cos^2(x) = \tfrac{1}{2}(1 + \cos(2x))$, we obtain a discrete Fourier expansion:
\begin{equation}
C(\alpha,\theta) = 2^{-N} \sum_{\epsilon \in \{0,1\}^N} \cos\!\left(2(\alpha-1)\theta \sum_{n=1}^N \epsilon_n \alpha^{-n} \right),
\end{equation}
which converges uniformly for each fixed $\theta$ as $N \to \infty$.
The frequencies $k = 2(\alpha-1)\sum_n \epsilon_n \alpha^{-n}$ represent spectral components of the Bernoulli convolution measure.
When $\alpha$ is Pisot, these frequencies cluster near integers, producing quasiperiodic, non-decaying oscillations.

The asymptotic behavior depends on the Diophantine properties of $\alpha$.
If $1/\lambda = \alpha$ is a Pisot number (an algebraic integer $>1$ whose conjugates all lie strictly inside the unit circle), then by a classical theorem of Erd\H{o}s \cite{Erdos1939,Erdos1940},
\(
\limsup_{\theta \to \infty} C(\alpha,\theta) > 0
\)
from the equivalence established above.
Intuitively, powers of a Pisot number are almost integers modulo $2\pi$, so for infinitely many $\theta$, the sequence $\{ \lambda^n (\alpha - 1)\theta \}$ clusters near zero, keeping most cosine factors near $+1$ and preventing decay.
If $1/\lambda = \alpha$ is not Pisot, then $\widehat{\nu}_\lambda(t) \to 0$ as $|t| \to \infty$.
Solomyak \cite{Solomyak1995,PeresSolomyak1996} proved that for almost all $\lambda \in (1/2,1)$, $\nu_\lambda$ is absolutely continuous with an $L^2$ density, implying $\widehat{\nu}_\lambda(t)$ is square-integrable and vanishes at infinity. Consequently,
\(
C(\alpha,\theta) \to 0.
\)
For non-Pisot $\alpha$, both numerical and analytical results \cite{PeresSolomyak1996} show power law decay with exponent at most $2$:
\begin{equation}
C(\alpha,\theta) = O(|\theta|^{-\gamma(\alpha)}),\quad 0<\gamma(\alpha)\le2,
\end{equation}
where the exponent $\gamma(\alpha)$ depends continuously on $\alpha$.
We note the special case $\alpha = 2$ saturates the bound with $\gamma(\alpha)=2$.  This explains the good performance that we see in numerical tests for a finite number of cycles with fixed total time.
When $\alpha=2$, $\lambda = 1/2$, giving the exact identity
\(
\prod_{n=1}^\infty \cos(\theta/2^n) = \frac{\sin \theta}{\theta},
\)
so that
\begin{equation}
C(2,\theta) = \left( \frac{\sin \theta}{\theta} \right)^2 \sim \theta^{-2}.
\end{equation}

The arithmetic dependence of $C(\alpha,\theta)$ reveals links between number theory and harmonic analysis.
Pisot parameters yield self-similar measures with non-decaying Fourier transforms, while non-Pisot parameters lead to equidistribution and decay.
Thus $C(\alpha,\theta)$ encodes subtle Diophantine structure in a deceptively simple cosine product.
The power-law decay of the product function also demonstrates that a single superiteration with fixed $\alpha$ is slow, while dividing the total time across multiple superiterations will give exponential suppression for large $T.$
Physically, as $T$ grows, a denser ``comb" of frequencies is required to suppress the increasingly fine features of the spectral function; driving $\alpha \to 1$ flattens the geometric series to achieve this, populating the frequency domain more densely to avoid the resonances seen with Pisot numbers.

As noted above, although the product function establishes the performance for a given $\theta$ and thus $E$, it establishes an upper bound on the integrated performance. Hence, for non-Pisot fixed $\alpha$ we have shown $\zeta = \Omega (\theta^{-\gamma (\alpha)})$ while for $\alpha$ that decreases with $\theta$ as $\alpha = 1 +b\theta^{-1},$ $\zeta$ is exponentially suppressed. We have hence constructed a set of times that asymptotically perform with exponential suppression in the optimization space of generalized superiterations.

We compare to the Gaussian-random RA (RRA) which samples times from the positive half of a Gaussian distribution with RMS value for $t$ being $\sigma$.
The average success probability for one cycle is
\begin{equation}
  \frac{1}{\sqrt{2\pi}\sigma}\int \mathrm dt\, e^{-\frac{t^2}{2\sigma^2}} \cos^2 [(E_{k}-E)\tfrac{t}{2}] =  \frac{1+e^{-(E_{k}-E)^2\sigma^2/2} }{2}.
\end{equation}
The success probability for $N$ cycles is the quantity above to the $N$-th power. But after $N$ time samples, we have an average total time of $T = N\sigma \sqrt{2/\pi}$, so we write the average success probability for $N$ cycles as
\begin{equation}
 \exp \left\{\frac{T}{\sigma}\sqrt{\frac{\pi}{2}}\log \left[ \frac{1+e^{-(E_{k}-E)^2\sigma^2/2} }{2} \right]\right\}.
\end{equation}
We see that we get exponential suppression for the Gaussian random rodeo algorithm with fixed $\sigma$.

Although the RRA provides exponential suppression on average, the shot-by-shot performance of the RRA drastically fluctuates and one may often get sub-exponential performance. Contrastingly, by using the generalized superiterations in the form above, we remain in the exponential suppression regime shot-by-shot.

\section{Performance of Generalized Superiterations on Physical Hamiltonians}
\label{sec:physicalham}

To assess the practical applicability of the generalized superiteration scheme, we numerically investigate the performance for several representative physical Hamiltonians. We begin by considering the spin-$\tfrac12$ XX model, defined by the Hamiltonian
\begin{equation}
    H = J\sum _{\langle i, j\rangle}  \left( \sigma ^+_i \sigma ^- _j + \sigma ^-_i \sigma^+_j\right)
    \label{eq:xxham}
\end{equation}
where $\langle i,j\rangle$ denotes nearest-neighbor sites and $\sigma ^\pm _i = (\sigma ^x _i \pm i \sigma ^y _i)/2 $ are the spin raising and lowering operators acting on site $i = 1,\ldots,L$. We consider the task of preparing the ground state of the zero-magnetization sector where $\braket{S^z}=\left\langle \sum _i \sigma ^z _i \right\rangle = 0$. Initial states with well-defined $S^z$ expectation are conserved under the dynamics, since $[S^z, H] = 0$.

\begin{figure}[t!]
    \centering
    \includegraphics[width=\linewidth]{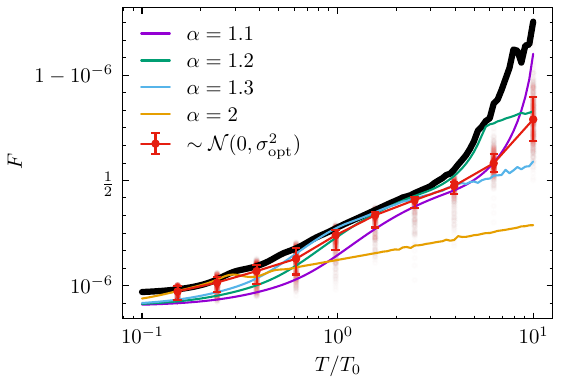}
    \caption{XX chain results for $\ket{\psi} = \ket{e_1}$ with $N=100$ time samples and a system size of $L=10$. We define the fidelity $F = |\braket{\psi '|\psi _0}|^2$  where $\ket {\psi _0}$ is the true zero-magnetization sector ground state. Solid colored lines represent fixed-$\alpha$ generalized superiterations that are fit to have total time $T$. The solid black line (Adaptive $\alpha$) corresponds to optimizing over $\alpha$ at each value of $T$. The red line with discrete markers corresponds to the averaged performance of the Gaussian random RA, with the distribution's variance optimized for each point.}
    \label{fig:xx_generalized_superiterations}
\end{figure}

\begin{figure}[t!]
    \centering
    \includegraphics[width=\linewidth]{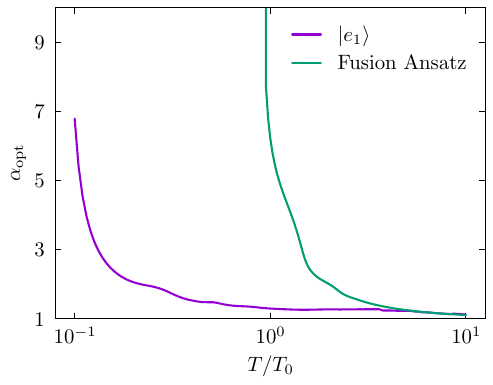}
    \caption{Optimal $\alpha$ across $T$ for the XX chain for the RA ansatz $\ket {e_1}$ and the fusion ansatz.}
    \label{fig:optimalalphasxx}
\end{figure}

As a first example, we take an XX model with $L=10$ and take $\ket \psi = \ket{e_1} = \ket{\uparrow \uparrow \uparrow \uparrow \downarrow \uparrow \downarrow \downarrow \downarrow \downarrow}$, i.e., the second basis vector in the ordered zero-magnetization subspace. This choice constitutes a deliberately poor initial guess, with an initial overlap with the ground state of only $|\braket{e_1 | \psi _0}|^2\approx 7\times 10^{-8}.$ This setup relaxes the assumption of a constant spectral overlap, and by considering a finite physical system, we also relax the assumption of a strictly continuous excitation band between $\Delta _{\text{min}}$ and $\Delta _{\text{max}} $.
We present the results for the fidelity as a function of total time $T$ in Fig.~\ref{fig:xx_generalized_superiterations}. We observe that fixing a single value of $\alpha $ across all total evolution times $T$ is inefficient and results in sub-optimal performance. For short total times $T \ll T_0$, superiterations with $\alpha = 2$ outperform those with smaller values of $\alpha$. For large $T$, however, fixed-$\alpha$ superiterations slow to a power-law increase in fidelity, in agreement with the analytical predictions of the preceding section. Contrastingly, the performance while adaptively optimizing $\alpha$ results in the thick black curve.
From the purple line in Fig.~\ref{fig:optimalalphasxx} we see that decreasing $\alpha$ toward $1$ as the allowed time $T$ is increased results in exponential performance. Finally, the average random-RA performance demonstrates its exponential suppression; however, especially for large times $T\gg T_0$, the generalized superiteration scheme with optimized $\alpha$ greatly outperforms the average RRA. Moreover, the shot-by-shot RRA performance (denoted by light red circles) demonstrates the large variance in RRA performance, especially for $T\gg T_0$, while the optimized generalized superiterations exhibit substantially reduced shot-to-shot variance.

\begin{figure}[t!]
    \centering
    \includegraphics[width=\linewidth]{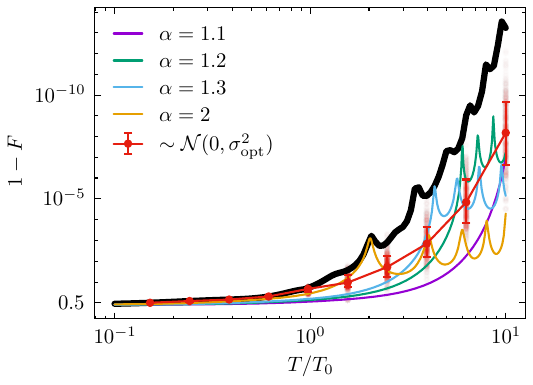}
    \caption{XX chain with $N=100$ time samples and a system size of $L=10$. We begin with a fusion ansatz. Minimization over $\alpha$ is done assuming it is monotonically decreasing.}
    \label{fig:xx_generalized_superiterations_fusion}
\end{figure}

As a further benchmark, we consider the input state $\ket \psi$ being the tensor product of the ground state of two disjoint 5-qubit XX Hamiltonians. As studied in \cite{patkowski2025high}, such an ansatz will provide better performance and a larger system may be constructed in a modular fashion by subdividing the system. Consequently, this choice of input state represents a more practical and scalable option for applications of the RA. The initial fidelity with the ground state is approximately $0.5$. As seen in Fig.~\ref{fig:xx_generalized_superiterations_fusion}, the choice of $\alpha =2$ superiterations performs near-optimally for a larger range of $T$ values, while it and other fixed-$\alpha$ approaches asymptote to slow power-law as $T$ is increased past a few $T_0$. Additionally, the RRA performance also remains exponential but slower than the optimal-$\alpha$ approach.
From the green line in Fig.~\ref{fig:optimalalphasxx}, we see that the optimal $\alpha$ similarly goes to one for large $T$ while it diverges at a greater value of $T$ compared to the other studied choice of input RA state.

\begin{figure*}[t!]
    \centering
    \includegraphics[width=\linewidth]{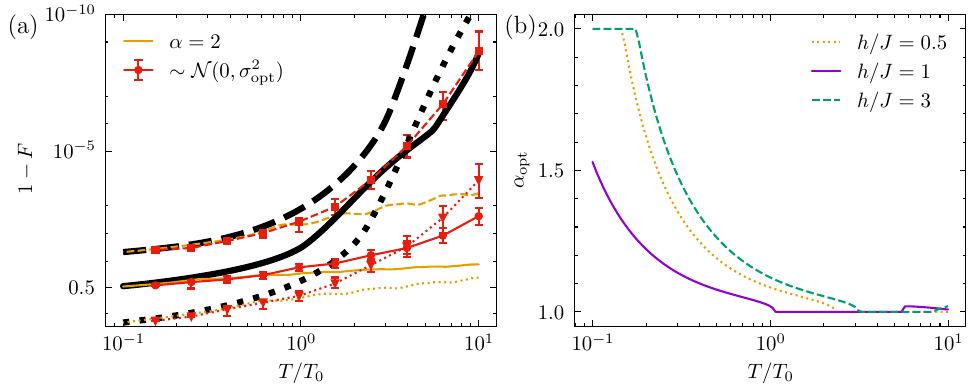}
    \caption{TFIM results. (a) Fidelity versus total time for various values of $h/J$. All solid lines indicate $h/J=1$, dotted lines correspond to $h/J=  0.5$, and dashed lines correspond to $h/J = 3$, as indicated by the legend of (b). We plot the superiterations with $\alpha = 2$ for reference in the golden lines. The red datapoints indicate the RRA performance while the black lines correspond to the optimal generalized superiteration performance. (b) Value of the optimal $\alpha$ as a function of total time. $\alpha$ is capped at $2$, hence the plateau at small $T$ where the true optimum is greater than 2.}
    \label{fig:TFIM_RA_combined}
\end{figure*}
As a further benchmark, we consider the one-dimensional transverse-field Ising model (TFIM),
\begin{equation}
    H = -J\sum _{i=1}^L \sigma ^z _i \sigma ^z _{i+1}-h \sum _{i=1}^L \sigma ^x _i,
\end{equation}
where $i=L+1 = 1$ corresponds to periodic boundary conditions.
The TFIM provides a complementary test case to the XX model as 
varying the transverse field $h$ tunes the system across the Ising critical point at $h/J = 1$ where the gap closes in the thermodynamic limit, constituting a particularly challenging case for the RA where the characteristic time diverges. In the finite-system-size limit, however, the gap remains finite but is minimized across this critical point. This tunability allows us to test the RA under both gapped and near-critical conditions. We note that the TFIM possesses a parity symmetry $P = \prod _i \sigma ^x _i$. We hence start with a simple-to-prepare state being the product state of the $\sigma ^x _i \ket + = \ket +$ state on all qubits and project it into the $+1$ eigenvalue subspace of $P$ such that it is an even parity state and use it as the input state $\ket \psi$. We similarly project the above $H$ into the even-parity sector for computational speedup, although starting with an input state satisfying $P\ket \psi = +1 \ket \psi$ guarantees we conserve the $\mathbb Z_2$ parity under dynamics with $H$ and with $R$, the RA propagator, since $[H,P] = 0.$

See Fig.~\ref{fig:TFIM_RA_combined} for the TFIM results. Fig.~\ref{fig:TFIM_RA_combined}(a) demonstrates that the optimally tuned geometric superiteration outperforms all alternatives by many orders of magnitude across the Ising phase transition. As shown in Fig.~\ref{fig:TFIM_RA_combined}(b), near the critical point $h/J \approx 1 $ the optimal value of $\alpha$ approaches unity significantly more rapidly than in noncritical regimes.

\section{Conclusion}
The original Rodeo Algorithm \cite{Choi:2020pdg} samples times from a Gaussian distribution $t_i \sim \left|\mathcal N(0, \pi\Delta^{-1}_{\text{min}})\right|$ which allows for sufficient average performance; however, the shot-to-shot performance is random and may perform sub-optimally.
As shown in Ref.~\cite{cohen2023optimizing}, a geometric series of times with a common ratio of $1/2$ will provide better performance on average. In this work, we have demonstrated that relaxing the common ratio assumption to a general $1/\alpha$ given sufficient resources allows one to perform even better when optimized over the value of $\alpha$, and empirical evidence shows this solution, termed \textit{generalized superiterations}, is near-optimal.

We further show that while generalized superiterations are a favorable optimization subspace, the parameter $\alpha$ must be adapted to system-specific parameters, otherwise the suppression is power-law at best. Analytics show that decreasing $\alpha$ as $\mathcal O(\theta ^{-1})$ with $\theta \sim (E-E_t)t$ provides exponential suppression of the eigenstate $\ket E$, and the random-RA similarly has, on average, exponential suppression.
We observe for two physical models, the XX chain with open BCs and the TFIM with periodic BCs, consistent qualitative behavior of the optimal $\alpha $ going to one for $T \gg T_0$ while for $T\lesssim T_0$ the optimal $\alpha$ increased beyond 2. This behavior persists across choices of input states as well as for various models where the excitation gap may be controlled. In particular, the optimized generalized superiteration scheme is robust across the TFIM phase transition.

We emphasize that although generalized superiterations do not achieve optimal performance, we empirically see that they are near-optimal and have exponential asymptotics. We conclude that for all practical purposes, optimizing over the value of the common ratio will provide one with a well-performing set of times that is easily attainable without complex or highly system-dependent optimization procedures. We also note that the value of $\alpha$ need not be fine-tuned and a plateau of acceptable performance exists around the optimal value, suggesting robustness to small errors in time sampling, such as those arising from experimental imperfections or Trotterization errors.

\begin{acknowledgements}
   This research is supported in part by U.S. Department of Energy (DOE) Office of Science grants DE-SC0023658, DE-SC0024586, DE-SC0026198, and DE-SC0013365 and U.S. National Science Foundation (NSF) grant PHY-2310620. This work has also been supported in part by NSF grant CHE-2154028 to KLCH.
\end{acknowledgements}

\clearpage
\onecolumngrid

\appendix
\section{Suppression for Continuous Band of Energies}
\label{app:suppressionderivation}

Plugging in the expression for $\xi '(E)$ from Eq.~\eqref{eq:postrastateexpressionineigenbasisofH} into the expression for the RSN (Eq.~\eqref{eq:residualspectralnorm}), and using the assumptions described in sec.~\ref{sec:modelham} we obtain:
\begin{equation*}
    \zeta = \int _{\Delta _{\text{min}}} ^{\Delta _{\text{max}}} \mathrm d E\cdot  \frac{1}{2^{2N}} \cdot |\xi (E) |^2 \prod _{n=1}^N\left| 1+ \exp [i(E_t -E)t_n] \right|^2.
\end{equation*}
Given a constant spectral norm that is properly normalized such that $|\xi (E)|^2 = (\Delta _{\text{max}} - \Delta _{\text{min}})^{-1}$ and taking $E_t=0$, we may rewrite the RSN as
\begin{equation}
\label{eq:unsimplifiedzeta1}
    \zeta = \frac{(\Delta _{\text{max}} - \Delta_{\text{min}})^{-1}}{2^{2N}}\int _{\Delta _{\text{min}}}^{\Delta _{\text{max}}} \mathrm dE \prod _{n=1}^N\left| 1+ \exp (-i Et_n) \right|^2.
\end{equation}
Now define
\begin{equation}
\label{eq:integralIdef}
    I(\Delta) = \int _{-\Delta}^\Delta \mathrm d E \prod _{n=1}^N \left|1+\exp \left(-iEt_n\right)\right|^2,
\end{equation} and recognizing that our integrand in Eq.~\eqref{eq:unsimplifiedzeta1} is even under $E\to -E,$ we may write
\begin{equation}
\label{eq:unsimplifiedzeta2}
\zeta = \frac{(\Delta _{\text{max}} - \Delta_{\text{min}})^{-1}}{2^{2N}} \cdot \frac 12 \left(I(\Delta _{\text{max}} ) - I(\Delta _{\text{min}})\right).
\end{equation}

Now we must simplify the $I$ integral from Eq.~\eqref{eq:integralIdef}. Note that the argument of the product may be written as
\begin{equation*}
    2 + \exp (-iEt_n) + \exp (iEt_n) = \sum _{k_n ,k_n'
     = \pm 1} \exp \left[iEt_n \cdot \frac{k_n + k_n '}2\right],
\end{equation*}
so Eq.~\eqref{eq:integralIdef} simplifies as:
\begin{align*}
    I(\Delta) &= \int _{-\Delta}^\Delta \mathrm dE \prod _{n=1}^N \sum _{k_n ,k_n'
     = \pm 1} \exp \left[iEt_n \cdot \frac{k_n + k_n '}2\right] \\
     &=\sum _{k_1 , k_1 ' = \pm 1} \cdots \sum _{k_N, k_N' = \pm 1} \int _{-\Delta}^\Delta\mathrm dE  \prod _{n=1}^N    \exp \left[iEt_n \cdot \frac{k_n + k_n '}2\right] \\
     &=\sum _{k_1 , k_1 ' = \pm 1} \cdots \sum _{k_N, k_N' = \pm 1} \int _{-\Delta}^\Delta\mathrm dE    \exp \left[\frac{iE}2 \sum _{n=1}^N  t_n (k_n + k_n ')\right] \\
     &= 2\Delta \sum _{k_1 , k_1 ' = \pm 1} \cdots \sum _{k_N, k_N' = \pm 1} \mathrm{sinc }\, \left[\frac \Delta 2 \sum _{n=1}^N  t_n (k_n + k_n ')\right].
\end{align*}
Plugging back in to Eq.~\eqref{eq:unsimplifiedzeta2}, we obtain:
\begin{align}
\label{eq:rsn_constant_overlap_continuous_spectrum}
    \zeta &= \frac{(\Delta _{\text{max}} - \Delta_{\text{min}})^{-1}}{2^{2N}} \cdot \Delta _{\text{max}}  \sum _{k_1 , k_1 ' = \pm 1} \cdots \sum _{k_N, k_N' = \pm 1} \mathrm{sinc }\, \left[\frac {\Delta _{\text{max}}} 2 \sum _{n=1}^N  t_n (k_n + k_n ')\right]\nonumber\\
    &- \frac{(\Delta _{\text{max}} - \Delta_{\text{min}})^{-1}}{2^{2N}}\cdot \Delta _{\text{min}} \sum _{k_1 , k_1 ' = \pm 1} \cdots \sum _{k_N, k_N' = \pm 1} \mathrm{sinc }\, \left[\frac {\Delta _{\text{min}}} 2 \sum _{n=1}^N  t_n (k_n + k_n ')\right].
\end{align}

We can also gain insight from the above approach for a superiteration with $\alpha = 2$, making the set of times $t_n = t_1 \cdot 2^{-(n-1)}$.
Such choice of times simplifies the expression of Eq.~\eqref{eq:unsimplifiedzeta1} in the limit $N\to \infty$ to read:
\begin{equation}\zeta = \frac 1 {(\Delta _{\mathrm {max}} - \Delta _{\mathrm{min}})} \int _{\Delta _{\mathrm{min}}} ^{\Delta _{\mathrm{max}}} \mathrm dE \, \mathrm{sinc} ^2 \, \left(E t_1\right),\end{equation}
and integration gives a lengthy expression with the sine integral function.
In the limit $t_1 \gg \Delta _{\mathrm{max}}, \Delta _{\mathrm{min}}$, the RSN has the form, without loss of generality, taking $\Delta _{\mathrm{max}}, \Delta _{\mathrm{min}} > 0$,
\begin{equation}
  \zeta = \frac{1}{2t_1^2} \left(\frac{1}{\Delta _{\mathrm{min}}} - \frac 1 {\Delta _{\mathrm{max}}}\right) + \mathcal O \left(t_1^{-3}\right),
  \label{eq:superiterationRSN_Ninf_longtime}
\end{equation}
demonstrating the power-law convergence of the $\alpha = 2$ superiterations.

\section{Optimization Codes for Generalized Superiterations}
We provide a set of codes for optimizing the time sampling for a problem over the superiteration space at \cite{PatkowskiRODEOCode}. To use the code, the user inputs the target energy, the total allotted time, and the trotter step size. One also inputs the estimated initial spectral function $|\xi (E)|^2$. The code then proceeds to fit the optimal geometric series ratio $\alpha^{-1}$ fitting the given parameters, rounded down to the nearest multiples of the Trotter step size, that minimizes the RSN.

\begin{figure}[h!]
    \centering
    \includegraphics[width=0.75\linewidth]{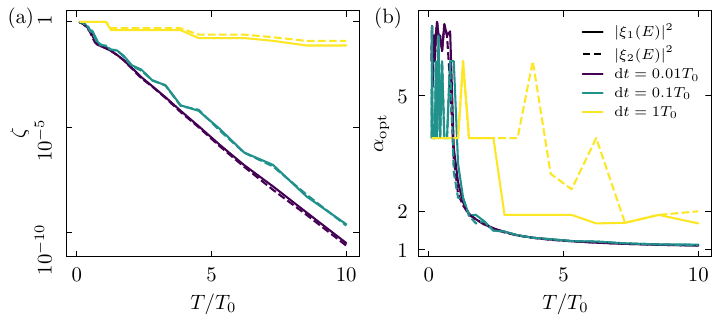}
    \caption{(a) RSN and (b) optimal $\alpha$ as a function of the total allotted time for various $\mathrm dt$. $|\xi_1 (E) |^2\propto e^{-E^2}$ and $|\xi_2(E)|^2\propto 1$ and both are zero outside of $E\in [0,1]$. $E_t$ is set to $-1$ such that $T_0 = \pi$.}
    \label{fig:opt_alphas_code}
\end{figure}

As an example application, we show below the results for the following setup: $E_t = -1$, $T\in [10^{-1} T_0,10 T_0]$, and Trotter step size $\mathrm dt \in \{10^{-2}T_0, 10^{-1}T_0, T_0\}$. Two spectral functions are studied. The first is a spectral function with most overlap near the lowest excited states, given as (up to normalization) $|\xi _1(E)|^2 = e^{-E^2} $ for $0\leq E \leq 1 $ and $0$ otherwise. The other is the constant spectral function $|\xi _2 (E)|^2= 1$ for $0\leq E \leq 1$ and $0$ otherwise. The results for the RSN and optimal $\alpha$ are shown in Fig.~\ref{fig:opt_alphas_code}.

\bibliography{References}

\end{document}